\documentclass[twocolumn,showpacs,preprintnumbers,amsmath,amssymb,prl]{revtex4}
\usepackage{amssymb}
\usepackage{amsmath}
\usepackage{hyperref}
\usepackage{mathrsfs}
\usepackage{graphicx}
\usepackage[utf8]{inputenc}
\usepackage{tipa}
\usepackage{color}
 \begin{document}
 \title{The Ericsson Nano-Brownian Engine in the Quantum Domain}
\author{S. Dattagupta}
\email{sushantad@gmail.com} 
\affiliation{Bose Institute, Kolkata 700054, India}                           
\author{S. Chaturvedi}
\email{subhash@iiserb.ac.in} 
\affiliation{Indian Institute of Science Education and Research, Bhopal 462066, India }
\date{\today}
\begin{abstract}
We examine here a hitherto unchartered Ericsson motor, operating in the quantum domain. The engine is a nanoscopic system of an electron trapped in a two-dimensional parabolic well and further subjected to an external magnetic field in the third direction. The quantum-coherent cyclotron motion of the electron is de-cohered due to strong interaction with a dissipative quantum heat bath.The calculation employs two different approaches – exact functional integral representation of the partition function and quantum Langevin equations for the operators of the system. Though in equilibrium, both the approaches yield equivalent expressions for the efficiency, the Langevin method opens up further avenues for investigating time dependent properties of the nano motor.
\end{abstract}
\pacs{03.65.Yz,05.30.-d,05.40.-a,05.70.Ln}
\maketitle

Nineteenth century witnessed the advent of a plethora of engines – some theoretical constructs, some practical implements – 
to take forward the industrial revolution. The objective in all these efforts is the same – 
to convert heat into mechanical work. Though the most talked about engine is perhaps what follows an ideal Carnot cycle (1824), 
the forerunner is the Stirling engine (1816). Then comes a succession of improved concepts – in the form of Otto-, Rankine-, and finally, 
the Ericsson-cycle (1853). Concomitant with this endeavour is the development of thermodynamics in the form of its first two laws which provide 
limits of efficiency that these engines can be pushed to. 

The subject of thermodynamics itself has seen important advances in the last few decades. 
Ordinarily, thermodynamics deal with systems in thermal equilibrium. But in recent years the subject has evolved in order to incorporate non-equilibrium 
and time-dependent phenomena. The latter often manifest as dissipative effects which are an inevitable consequence of the system of interest being in 
contact with the environment that acts as a heat bath, giving rise to a novel subject called Stochastic Thermodynamics \cite{1}-\cite{4}. 
 When the system-size is small and the temperature is low, quantum mechanics naturally come into play in 
dealing with these systems \cite{5},\cite{6} and the said effects become particularly relevant for the contemporary challenging topic of nano materials 
which finds numerous realizations in condensed matter physics and biology. This has prompted the necessity of combining thermodynamics of 
dissipative systems with quantum mechanics, particularly for what are called open systems \cite{7}-\cite{8a}. Hence, we are led to studying what may 
be viewed as the quantum version of stochastic thermodynamics, the latter having been recently implemented in the laboratory for a micrometer-sized biological motor operating between bacterial reservoirs \cite{9}.

Normally, quantum dissipation is studied when the system of interest is weakly coupled with the environment, as in the familiar linear response 
theory or master equation method that abound in condensed matter physics and quantum optics \cite{7}. In such situations the steady-state 
(i.e. time-independent) 
thermodynamic properties do not depend on dissipative parameters characterizing friction, damping, diffusion coefficient, etc., 
the latter being attributes of the heat bath. 

The last three decades have however seen an upsurge of activity in the strongly coupled regimes, wherein it is not at all sensible to delineate 
what is a system and what is a heat bath \cite{8},\cite{10}-\cite{13}. All that one can reasonably do is to identify the relevant degrees of freedom – most often 
dictated by experiments – to construe what is called a system of interest. The rest of the degrees of freedom, belonging to what may be viewed 
as the environment, are averaged out or integrated out, as the case may be, maintaining though the condition of strong coupling. 
To give an example, consider an experiment on nuclear magnetic resonance. The relevant degrees of freedom that couple to laboratory 
magnetic fields belong to the nuclei, whereas the environment comprises other degrees of freedom such as those of electrons and phonons. 
Interestingly, the theory of strongly coupled dissipative quantum mechanics of open systems has further elucidated the third law of thermodynamics 
\cite{11}-\cite{13}. 

Our aim in this paper is to return to the issue of engines, mentioned in the introductory paragraph,  
in the light of all these recent developments in nano science and strongly dissipative open quantum systems \cite{11}-\cite{13}. 
 In recent years, several microscopic realizations of the Otto, Carnot and Stirling engines employing varying `working 
substances' have been considered in the literature both from theoretical as well as experimental points of view \cite{9},\cite{14}-\cite{21}
with particular attention to questions pertaining to their efficiencies.
Our focus is a Brownian motor – of great significance in living systems – which is a magnetic version of the Ericsson cycle that, 
interestingly, resembles the first in the chain of engines, viz., the Stirling cycle considered in \cite{14} realized as a  
a nano Brownian motor in the form of a quantum harmonic oscillator that 
undergoes quantum diffusive motion \cite{14},\cite{22}. Our nano system of interest, on the other hand, is an electron that can be trapped in a 
two-dimensional parabolic well whose orbital motion is affected by a laboratory magnetic field $B$ \cite{23}-\cite{26} . Thus, we are led to consider
a ‘magnetic Ericsson cycle’ in which the pressure $P$ is replaced by -$B$ and the volume $V$ by the magnetization $M$. The cycle consists of 
an isothermal ‘compression’ (accompanied by a decrease in $M$), followed by an ‘isobaric’ (corresponding to a constant $B$-field) heating, 
then an isothermal ‘expansion’ (accompanied by an increase in $M$), and finally, an ‘isobaric’ cooling, as shown schematically below.
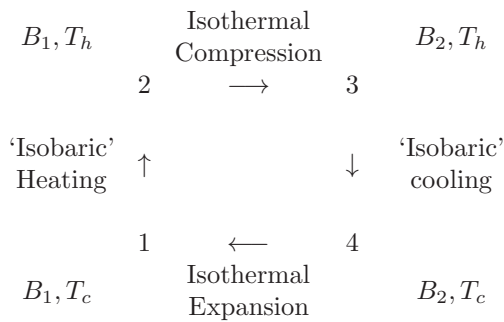
\begin{align}
&\begin{array}{cccccc}
B_1,T_h~&  &\begin{array}{c} \text{Isothermal}\\ \text{Compression}\end{array}& &&B_2,T_h\\
& 2 &\longrightarrow& 3&&\\
&  && &&\\
\begin{array}{c}\text{`Isobaric'}\\ \text{Heating}\end{array}&\uparrow&&\downarrow&&\begin{array}{c}\text{`Isobaric'}\\\text{cooling}\end{array}\\
&  && &&\\
& 1&\longleftarrow&4&&\\
B_1,T_c~& &\begin{array}{c}\text{Isothermal}\\ \text{Expansion}\end{array}& &&B_2,T_c\\
&&&&&\\ \end{array}
\nonumber
\end{align}

In the spirit of  \cite{14} our first task is to calculate the efficiency from the thermodynamic expressions for the Helmholtz free energy and the internal energy. 
To this end we employ the known results from an exact functional integral calculation of the partition function $Z$ of an electron in a parabolic well 
plus a magnetic field, coupled with an infinitely large number of quantum harmonic oscillators that mimic the environment \cite{27},\cite{28}. 
In contrast to \cite{14} 
which is based on a weak-coupling approximation our result for $Z$ depends on damping parameters, as expected. We now resort to quantum Langevin equations 
for time-dependent position and velocity operators of the electron \cite{23}-\cite{29} -- again exact -- and rewrite them in terms of ‘noisy’ work, energy and heat 
operators a la the first law of thermodynamics. The stratagem is a quantum adaptation of the stochastic thermodynamic scheme of Sekimoto and 
others \cite{1},\cite{2}.

As we said at the outset our main interest is in the computation of the efficiency $\eta$ of the Ericsson engine at all temperatures especially where quantum 
effects dominate.

The basic thermodynamic relation that applies when the laboratory magnetic field replaces the pressure \cite{30} is
\begin{equation}
TdS = dE – BdM = d(\varepsilon + BM) – BdM = d\varepsilon + MdB. 
\label{1}
 \end{equation}                                                 
The first law of thermodynamics then reads
\begin{equation}
dQ = d\varepsilon + dW = d\varepsilon + MdB.
\label{2}
\end{equation}
                                                                          
Note it is the energy $\varepsilon (= E – BM)$ – a Legendre transform to the ‘internal energy’ $E$ -- 
that connects to statistical mechanics in the sense that 
it equals the expectation value of the system Hamiltonian $\hat{H}_{S}$ which, in the present problem, reads 
\begin{align}
\hat{H}_S &= [(\hat{p}_x - eB\hat{y}/2c)^2 + (\hat{p}_y + eB\hat{x}/2c)^2]/2m\nonumber\\
&+ m\Omega_0^2 (\hat{x}^2 + \hat{y}^2)/2, 
\label{3}
\end{align}
where - $e$ is the charge of the electron of mass $m$, $p$’s are the canonical momenta along $x$ and $y$-axes and the last term represents the potential 
energy in the parabolic well. Here we have chosen to invoke the symmetric gauge of the vector potential associated with a $B$-field in the $z$-direction 
\cite{23}-\cite{29}. Accordingly, the Helmholtz free energy $F$ is \cite{30}
\begin{align}
F = - K_BT \text{ln} Z =    \varepsilon  - TS, 
\label{4}
\end{align}
$Z$ being the canonical partition function. From Eq. (1) then
\begin{align}
M =  - \left( \frac{\partial F}{\partial B}\right )_{T}, S = -  \left(\frac{\partial F}{\partial T}\right)_B. 
\label{5}
\end{align}
A combination of Eqs. $(\ref{2})$ and $(\ref{5})$ then yields for the work done in an isothermal process as the magnetic field is increased 
from $B_1$ to $B_2$ as
\begin{align}
W_{1\rightarrow 2} = F (T, B_2) – F (T, B_1).  
\label{6}
\end{align}
We now make the system described by Eq. $(\ref{3})$ an open one by expanding the Hamiltonian \cite{23}-\cite{29}: 
\begin{align}
\hat{H} = \hat{H}_S + \sum_{j}\left[  \frac{\hat{{\bf p}}_j^2}{2m_j} + \frac{1}{2}m_j \omega_j^2\left(\hat{{\bf q}}_j – C_j \frac{\hat{{\bf q}}}{ m_j 
\omega_j^2}\right)^2\right].  
\label{7}
\end{align}
The additional term, which contains a linear coupling between the system coordinate and the environment coordinate via a coupling constant $C_j$, 
would influence the dynamics of $\hat{H_S}$, and being representative of a very large system, would make this dynamics dissipative. By averaging over the 
additional degrees of freedom we can calculate the so-called reduced partition function from which the Helmholtz free energy and other 
thermodynamic potentials can be evaluated. We quote the known result for the   free energy $F$ \cite{27},\cite{28}:
\begin{align}
	F&=  -\frac{2}{\beta}\left(\frac{\beta\Omega_0}{4\pi^2}\right)-\frac{1}{\beta}\sum_{j=1}^{3}\left[\text{ln}\Gamma\left(\frac{\lambda_j}{\nu}\right)+     
\text{ln}\Gamma \left(\frac{\lambda_j^\prime}{\nu}\right)\right]\nonumber\\
	&+\frac{2}{\beta}\text{ln}\Gamma\left(\frac{\Omega_D}{\nu})\right) ,
 \label{8}
  \end{align}
  where $\nu$ is the bosonic Matsubara frequency(thermal energy devided by Planck constant) and   $\Gamma(z)$ is the Gamma function whose arguments are
\begin{align}
 \lambda_1+\lambda_2+\lambda_3&=\Omega_D+i\Omega_c\nonumber \\
 \lambda_1\lambda_2+\lambda_2\lambda_3+\lambda_3\lambda_1&=\Omega_0^2+\gamma\Omega_D+i\Omega_c\Omega_D\nonumber\\
 \lambda_1\lambda_2\lambda_3 =\Omega_0^2\Omega_D.
 \label{9}
 \end{align}
The corresponding primed $\lambda$'s are obtained by taking complex conjugates of those in $(\ref{9})$. In writing the above formulae we have 
assumed Ohmic dissipation with constant damping $\gamma$, a high-frequency Drude cut-off $\Omega_D$ and have introduced the cyclotron 
frequency $\Omega_c (= |e|B/mc)$ and $\beta = (K_B T)^{-1}$.Throughout we take the Planck constant to be unity. It can easily be checked that for $\Omega_D\longrightarrow\infty$, the limit in which we finally choose to work, $\lambda_3 \longrightarrow\Omega_D$. Additionally for $\gamma=0$, Eq.$(\ref{8})$ matches with 
the free energy for the `free' system sans the heat bath $F_0$:
\begin{align}
	F_0=-\frac{1}{\beta}\text{ln}\left[\text{cosech}\left(\frac{-i\beta\lambda_1^0}{2}\right) \cdot\text{cosech}
	\left(\frac{-i\beta\lambda_2^0}{2}\right)\right],
	\label{10}
\end{align}
where the superscripted $\lambda$'s ensue in the limit of $\Omega_D\longrightarrow\infty$ and $\gamma=0$.

The internal energy $\varepsilon$ can be derived from the free energy $F$ by  employing
the identity.
\begin{align}
 \varepsilon =\left( \dfrac{\partial}{\partial \beta} (\beta F)\right)_{V}.
 \label{11}
\end{align}

Referring to the Ericsson cycle and Eq. $(\ref{6})$ the net work done by the engine during the two isothermal processes is 
\begin{align}
&\Delta W = - [F (B_1, T_c) – F (B_2, T_c)] + [F(B_1,T_h) – F(B_2,T_h)].
\label{12}
\end{align}
On the other hand, the heat absorbed at the hot source $T_h$ is
\begin{align}
       \Delta Q &= [F(B_1,T_h) - F(B_2,T_h)] +  [(\varepsilon)_3 – (\varepsilon)_2] + [(\varepsilon)_2 – (\varepsilon)_1] \nonumber\\
       &= [F(B_1,T_h) – F(B_2,T_h)] + [\varepsilon(B_2,T_h) – \varepsilon(B_1,T_c)].
       \label{13}
       \end{align}
 The efficiency of the Ericsson engine is defined by 
 \begin{align}
  \eta=\dfrac{\Delta W}{\Delta Q}.
  \label{14}
 \end{align}
 While an exact expression for $\eta$ can be obtained by use of Eqs.$(\ref{8}-\ref{13})$, we may derive further insights into the results by 
 looking at the low and high temperature limits as discussed in \cite{28}

The method outlined above is based on what Kadanoff calls the Gibbs approach to statistical physics, as opposed to the Einstein approach in which 
the underlying Brownian motion is analysed in the time domain via a quantum Langevin equation \cite{31}. If the latter has an in-built 
fluctuation-dissipation theorem the steady-state (i.e., the limit of time $t\rightarrow \infty$) properties should 
replicate the results derived from the Gibbs approach. We have seen earlier that the problem of dissipative cyclotron motion of an electron 
in a confined potential, as studied here, throws up intriguing issues of various limiting procedures, such as $t$ going to infinity 
and $\Omega_0$ going to zero \cite{26}-\cite{28}. Thus it is interesting to re-examine the efficiency of the Ericsson cycle from an underlying 
Langevin equation, which further clarifies the role of the ‘heat operator’ in terms of the ‘quantum noise’ generated by the environment 

Based on the Hamiltonian in Eq.$(\ref{7})$ we have derived earlier the following quantum Langevin equations, which for Ohmic dissipation read
\begin{align}
&\hat{v}_x \equiv \dfrac{d}{dt}\hat{x}= \frac{1}{m} \hat{p}_x - \frac{1}{2}\Omega_0 \hat{y},\nonumber\\
 &\hat{v}_y\equiv \dfrac{d}{dt}\hat{y}= \frac{1}{m} \hat{p}_y + \frac{1}{2}\Omega_0 \hat{x},\nonumber\\
&\frac{d}{dt}\hat{p}_x= -\frac{1}{2}\Omega_c \hat{p}_y - m\left(\frac{\Omega_c^2}{4} + \Omega_0^2\right) \hat{x} -
m\gamma \hat{v}_x+ \hat{f}_x ,\nonumber\\
&\frac{d}{dt}\hat{p}_y= ~~  \frac{1}{2}\Omega_c \hat{p}_x - m \left(\frac{\Omega_c^2}{4} + \Omega_0^2\right)\hat{y} - m\gamma \hat{v}_y+ \hat{f}_y, 
\label{15}
\end{align} 
(Here and later, for the sake of brevity, we suppress the the time arguments of the operators with the understanding that all the operators are evaluated at the 
same time $t$). The correlations of the noise operators appearing in $(\ref{15})$ in the case of Ohmic dissipation are given by 
where  
\begin{align}
 &\langle\{\hat{f}_\mu(t), \hat{f}_\nu(t')\}\rangle = \nonumber\\
 &~~~~~~~~\delta_{\mu\nu} \frac{2\gamma}{\pi}\int_0^\infty d\omega
  \omega~\text{coth}(\beta\omega/2) \cos[\omega(t-t')],  \nonumber\\      
&\langle[\hat{f}_\mu(t), \hat{f}_\nu(t')]\rangle  =  \delta_{\mu\nu} \frac{2\gamma}{i\pi}\int_0^\infty 
d\omega \sin [\omega(t-t')].  
\label{16}
\end{align}

Note that the operators $\hat{v}_x(t)$ and $\hat{v}_y(t)$ do not commute with each other.

Allowing the $B$-field to be time-dependent, the time derivative of the system Hamiltonian $H_S$ can be written as
\begin{align}
\dfrac{d}{dt}\hat{H}_S& = \dfrac{m}{2} \dfrac{d}{dt} [(\hat{v}_x^2 + \hat{v}_y^2)
+ \Omega_0^2(\hat{x}^2 + \hat{y}^2)]\nonumber\\
&- \frac{e}{2c} (\hat{y}\hat{v}_x-\hat{x}\hat{v}_y). \dfrac{d}{dt}B. 
\label{17}
\end{align}
Hence, following Sekimoto, the first law of thermodynamics in $(\ref{2})$ can be written in the operator form as
\begin{equation}
d\hat{Q} = d\hat{\varepsilon} +d\hat{W},  
\label{18}
\end{equation}
where
\begin{align}
d\hat{W} = -\hat{M} dB,~
\hat{M} \equiv \frac{e}{2c} (\hat{y}\hat{v}_x-\hat{x}\hat{v}_y), 
\label{19}
\end{align}
$\hat{M}$ being the magnetic moment operator and 
\begin{align}
d\hat{\varepsilon} = d( m[\hat{v}_x^2 + \hat{v}_y^2] /2 + m\Omega_ 0^2[\hat{x}^2 + \hat{y}^2]/2), 
\label{20}
\end{align}
the differential energy operator, and
\begin{align}
                               d\hat{Q} =  (- m\gamma \hat{v}_x + \hat{f}_x) d\hat{x} + (- m\gamma \hat{v}_y + \hat{f}_ y) d\hat{y},  
                               \label{21}
                               \end{align}
the differential heat operator. 

In this representation the first law acquires a microscopic meaning and all the quantities in Eqs.$(\ref{17})$-$(\ref{20})$ allow for calculation of 
not just their averages but fluctuations and probability distributions as well, employing the statistics of the noise operators $\hat{f}(t)$ 
which is a Gaussian but a non-white process.
While the set of Eqs. $(\ref{15})$ can be solved exactly and the time-dependence of all relevant quantities can be determined our aim is to extract 
the asymptotic equilibrium properties. Following \cite{28} we find that $\varepsilon =  \lim_{t\to\infty} $  $\langle \hat{H}_S(t)\rangle$
is identical to that given by Eq. $(\ref{11})$, at low temperatures. Naturally, the efficiency $\eta$ matches with that given by the Gibbs approach 
derived from Eqs. $(\ref{12})$ and $(\ref{13})$.

Summarizing, we present in this Letter new and exact results for the efficiency of a nano motor in the quantum regime. 
The treatment goes beyond the much studied damped quantum harmonic oscillator in that two-dimensionality brings-in features 
which open up further scope for a re-look at the inter-connected energy-, heat,- and work-fluctuations in the context of the 
topically important fluctuation theorems \cite{32}. Furthermore, the efficiency of the engine is shown to depend on damping just as 
another fundamental attribute of thermodynamics like the heat capacity does, providing additional clarity on not just the first two 
laws but the third law as well. The results are given for both the Gibbs and Einstein approaches to statistical mechanics, 
thus unifying different routes that are normally followed in studying the thermodynamics of nano systems. 
In future work we will further analyze the time-dependent exact solutions of the Langevin equations $(\ref{15})$ and address the issues of 
the transient effects as well as the finite-time corrections to the asymptotic solutions.  

{\bf Acknowlegements}
SD is grateful to the Indian National Science Academy, New Delhi, for supporting his research through their scheme of ‘Senior Scientists’. He is also 
thankful to the Indian Institute of Science Education and Research, Bhopal, for its kind hospitality during the time this paper was 
completed, and to J.Kumar for useful inputs.

\end{document}